\documentclass[5p]{elsarticle}
\usepackage{amsmath}
\usepackage{graphicx}
\usepackage[small,bf,nooneline]{caption}
\biboptions{sort&compress}
\journal{Physics Letters B}
\newcommand{\q}[1]{``#1''}

\begin{document}

\begin{frontmatter}

\title{Wavelet signatures of $K$-splitting of the Isoscalar Giant Quadrupole Resonance in deformed nuclei from high-resolution (p,p$'$) scattering off  $^{146,148,150}$Nd}

\author[wits,biust]{C.O.~Kureba}
\author[tlabs]{Z.~Buthelezi}
\author[wits]{J.~Carter}
\author[witsgeo]{G.R.J.~Cooper}
\author[uct]{R.W.~Fearick}
\author[tlabs]{S.V.~F\"ortsch}
\author[wits]{M.~Jingo}
\author[jinr,dres]{W.~Kleinig}
\author[tud]{A.~Krugmann}
\author[tud]{A.M.~Krumbolz}
\author[cu]{J.~Kvasil}
\author[sun]{J.~Mabiala}
\author[tlabs]{J.P.~Mira}
\author[jinr]{V.O.~Nesterenko}
\author[tud]{P.~von~Neumann-Cosel\corref{cor1}}\ead{vnc@ikp.tu-darmstadt.de}
\author[tlabs]{R.~Neveling}
\author[tlabs,sun]{P.~Papka}
\author[erl]{P.-G.~Reinhard}
\author[tud]{A.~Richter}
\author[wits]{E.~Sideras-Haddad}
\author[tlabs]{F.D.~Smit}
\author[tlabs]{G.F.~Steyn}
\author[sun]{J.A.~Swartz}
\author[osaka]{A.~Tamii}
\author[wits]{I.T.~Usman}
\cortext[cor1]{Corresponding author}

\address[wits]{School of Physics, University of the Witwatersrand, Johannesburg 2050, South Africa}
\address[biust]{Department of Physics and Astronomy, Botswana International University of Science and Technology, P. Bag 16, Palapye, Botswana}
\address[tlabs]{iThemba Laboratory for Accelerator Based Sciences, PO Box 722, Somerset West 7129, South Africa}
\address[witsgeo]{School of Earth Sciences, University of the Witwatersrand, Johannesburg 2050, South Africa}
\address[uct]{Department of Physics, University of Cape Town, Rondebosch 7700, South Africa}
\address[jinr]{Laboratory of Theoretical Physics, Joint Institute for Nuclear Research, Dubna, Moscow region, 141980, Russia}
\address[dres]{Technische Universit\"at Dresden, Institut f\"ur Analysis, D-01062, Dresden, Germany}
\address[tud]{Institut f\"ur Kernphysik, Technische Universit\"at Darmstadt, D-64289, Darmstadt, Germany}
\address[cu]{Institute of Particle and Nuclear Physics, Charles University, CZ-18000, Prague, Czech Republic}
\address[sun]{Department of Physics, University of Stellenbosch, Matieland 7602, South Africa}
\address[erl]{Institut f\"ur Theoretische Physik II, Universit\"at Erlangen, D-91058, Erlangen, Germany}
\address[osaka]{Research Center for Nuclear Physics, Osaka University, Ibaraki, Osaka 560-0047, Japan}

\begin{abstract}

The phenomenon of fine structure of the Isoscalar Giant Quadrupole Resonance (ISGQR) has been studied with high energy-resolution proton inelastic scattering at iThemba LABS in the chain of stable even-mass Nd isotopes covering the transition from spherical to deformed ground states. 
A wavelet analysis of the background-subtracted spectra in the deformed $^{146,148,150}$Nd isotopes reveals characteristic scales in correspondence with scales obtained from a Skyrme RPA calculation using the SVmas10 parameterization.
A semblance analysis shows that these scales arise from the energy shift between the main fragments of the $K$ = 0, 1 and $K$ = 2 components.
 
\end{abstract}


\end{frontmatter}

\section{Introduction}\label{intro}

Nuclear giant resonances are a prime example of elementary excitation modes and important sources of information about nuclear collectivity. 
The decay of these resonances provides information on how a well-ordered collective excitation dissolves into a disordered motion of internal degrees of freedom in fermionic quantum many-body systems~\cite{Bor98, Ric05}.
A systematic experimental investigation of the fine structure of the Isoscalar Giant Quadrupole Resonance (ISGQR) at different shell closures over a wide mass range with high-resolution proton inelastic scattering experiments was conducted at the K600 magnetic spectrometer of iThemba LABS \cite{She04,She09, Usm11}.  
These studies showed that in medium-mass to heavy nuclei, experimentally observed energy scales quantitatively characterizing the fine structure could be related to those extracted from microscopic calculations involving coupling of the initial collective one particle-one hole (1p-1h) excitations to the background of two particle-two hole (2p-2h) states and, in particular, to low-lying surface vibrations. 
The characteristic energy scales were extracted applying wavelet analysis techniques \cite{She08}.

Fine structure has been established in recent work to appear as a general phenomenon in low-multipolarity nuclear giant resonances. 
This holds not only for electric modes like the ISGQR or the Isovector Giant Dipole Resonance (IVGDR)~\cite{Die01, Str00,Pol14} but also for the magnetic quadrupole (M2)~\cite{Neu99}, spin-flip M1~\cite{lan04,pet10} and Gamow-Teller (GT)~\cite{Kal06} resonances. 
It has also been demonstrated for the example of the ISGQR that fine structure occurs in all but the lightest nuclei. 
In this work we explore the role of the deformation degree of freedom by a study of the stable even-mass neodymium isotope chain ($^{142,144,146,148,150}$Nd) representing a transition from spherical to quadrupole-deformed nuclei. 

The properties of the fine structure provide important insight into the dominant damping mechanisms of nuclear giant resonances as explained in detail in Refs.~\cite{Har01,bbb98}. 
The resonance width $\Gamma$ is commonly assumed to contain three contributions, viz.\ a fragmentation of the initial 1p-1h excitations acting as doorway states (called Landau damping), direct particle emission from the 1p-1h excitations expressed by an escape width $\Gamma\!\!\uparrow$, and coupling to the more complex two particle-two hole (2p-2h) and finally many particle-many hole (np-nh) states leading to a spreading width $\Gamma\!\!\downarrow$ due to internal mixing. 
In this context, fine structure may be understood to result from the characteristic time scales (or energies) of the coupling steps. 
In a broader sense, such structure intermediate between the typical widths of compound states and that of the total resonance might also be viewed to arise from a \q{super-radiant} mechanism~\cite{Aue07} in analogy to the Dicke coherent state~\cite{Dic54} of a set of two-level atoms coupled through the common radiation field. 

The present Letter focuses on the impact of deformation on the fragmentation of the ISGQR and thereby on the scales of the fine structure.
In axially deformed nuclei, one expected effect is a splitting of ISGDR to the components 
corresponding to different $K$ quantum numbers, where $K$ stands for the projection of the total angular momentum on the respective axis \cite{Boh98}.
Such a $K$-splitting has been clearly observed for the case of the IVGDR \cite{Ber75}.
The total photo-absorption cross sections for the neodymium isotopes nuclei with 142 $\leq A \leq$ 150 from Ref.~\cite{Car71} exhibit, not only a broadening of the Lorentzian shape with increasing ground-state deformation, but also a transition from a  single Lorentzian shape in the semimagic $^{142}$Nd two a double-Lorentzian structure in the most  deformed nucleus $^{150}$Nd (although the latter is called into question by recent results \cite{don16}).

No such effect has been seen in inelastic alpha scattering studies of the ISGQR in the Nd \cite{Gar84} or the analog Sm \cite{You04} isotope chains.
A two-component fit for the $E$2 strength distribution of the Sm isotopes has been proposed by Itoh et al.~\cite{Ito03}. 
However, the extra component at higher excitation energies is not related to deformation, but represents a phenomenological description of the continuum strength beyond the main ISGQR peak. 
A comparison of the ISGQR in $^{142}$Nd and $^{150}$Nd in inelastic electron scattering indicates a broadening and shift of the centroid towards higher excitation energy for the latter, which might be seen as indirect evidence for a $K$-splitting \cite{Sch75}. 
However, the investigation of broad structures in inclusive electron scattering is hampered by the large background due to the radiative tail of elastic scattering.

Theoretical calculations predict splittings between the $K$ components which are typically smaller than the Full Width at Half Maximum (FWHM) of the ISGQR observed experimentally. 
It is thus not surprising that gross structure, i.e.\ the $E$2 strength distributions, deduced from different experiments do not show clear signatures of $K$-splitting.
Here, we demonstrate using wavelet analysis techniques that in contrast, the fine structure contains fingerprints of the main fragments of the different $K$ components and their relative shift (the $K$-splitting).   
Although data were taken for all stable even-even Nd isotopes, we restrict ourselves in the following to $^{146,148,150}$ Nd. 
Indeed, a wavelet analysis of $^{142,144}$Nd is interesting and will be part of a subsequent paper. 
However, our previous survey of fine structure of the ISGQR established coupling to low-lying phonons as a major source of wavelet scales in spherical/vibrational nuclei \cite{She04,She09}, while the present manuscript has a focus on the role of ground state deformation.

\section{Experiment} \label{sec:experiment}

\begin{figure}[bh]
\begin{center}
\includegraphics[width=\columnwidth,angle=0]{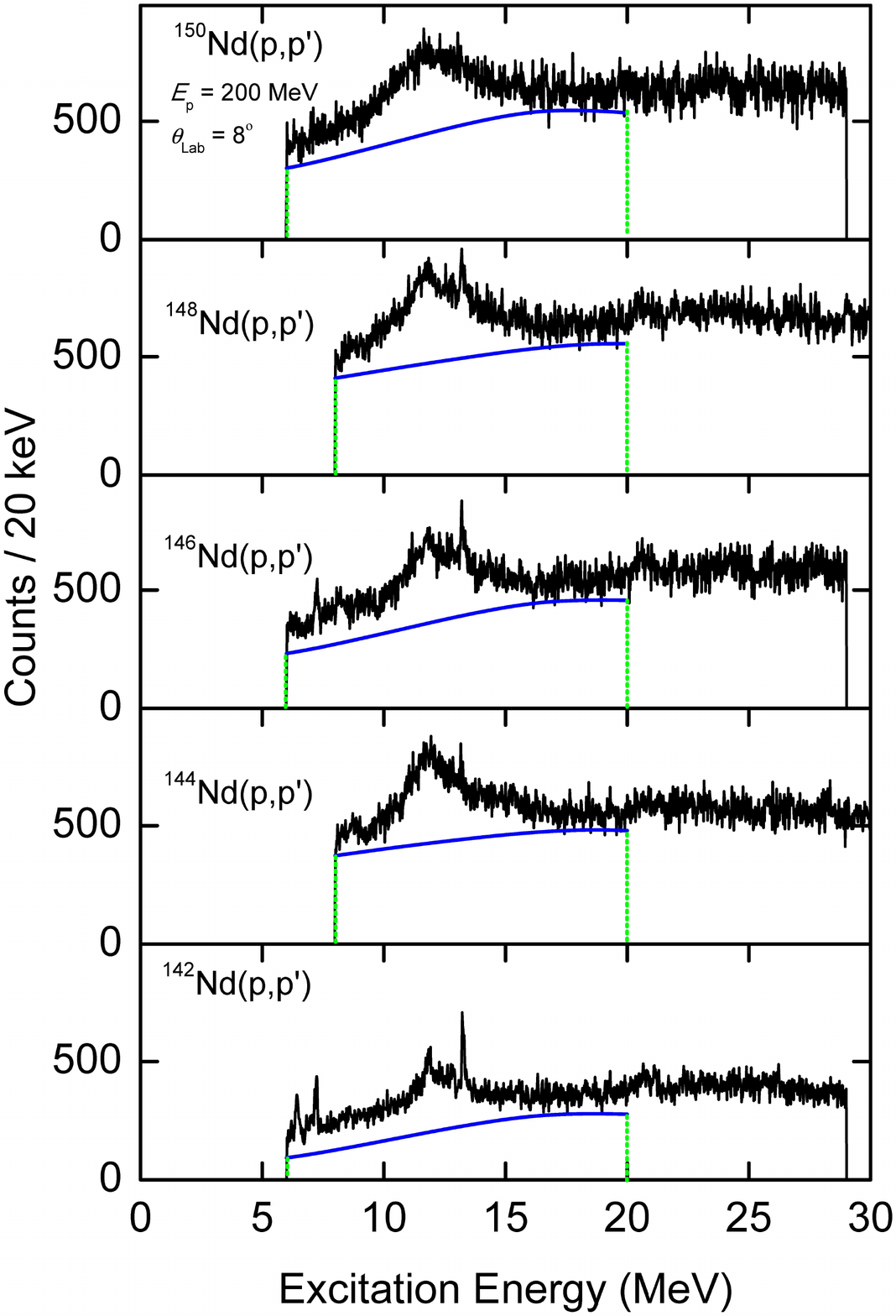}
\caption{Excitation energy spectra for proton inelastic scattering on stable even-mass neodymium isotopes at $E_{\rm p}$ = 200 MeV and $\theta_{\rm lab} = 8^\circ$.
The ISGQR is represented by the bump between about 9 and 15 MeV. 
The solid (blue) curves represent the levels of background from a DWT analysis and the dotted vertical (green) lines the energy range of application. 
The sharp peak around 13 MeV is a $^{16}$O contaminant.
}
\label{fig:inelastic-all}
\end{center}
\end{figure}
%
The experiment used {\it E}$_{\rm Lab}$ = 200 MeV proton beams from the Separated Sector Cyclotron facility of iThemba LABS. Measurements were made using the K600 magnetic spectrometer, where a high energy-resolution $\Delta E \approx 45$ keV FWHM could be achieved with dispersion matching techniques \cite{Nev11}.
Data were taken for highly enriched ($> 96\%$) $^{142,144,146,148,150}$Nd targets with $2 -3$ mg/cm$^2$ areal densities at a scattering angle $\theta_{\rm Lab}$ = 8$^{\circ}$, where the cross section of the ISGQR is at maximum. 
Details of the experiment and data analysis can be found in Ref.~\cite{Kur14}.
The resulting spectra covering an excitation energy range of about 8  to 28 MeV are shown in Fig.~\ref{fig:inelastic-all}.
The spectra contain some contamination from oxidation, e.g. the strong transition around 13 MeV visible in some of the targets. 
They were identified by a measurement using a mylar foil in the same kinematics. 
Since the peaks would modify all further analysis, they have been removed from the spectra and the corresponding energy region was replaced by a white spectrum with a variance determined by the statistics in neighbouring channels.

\begin{figure}[tbh]
\begin{center}
\includegraphics[width=\columnwidth,angle=0]{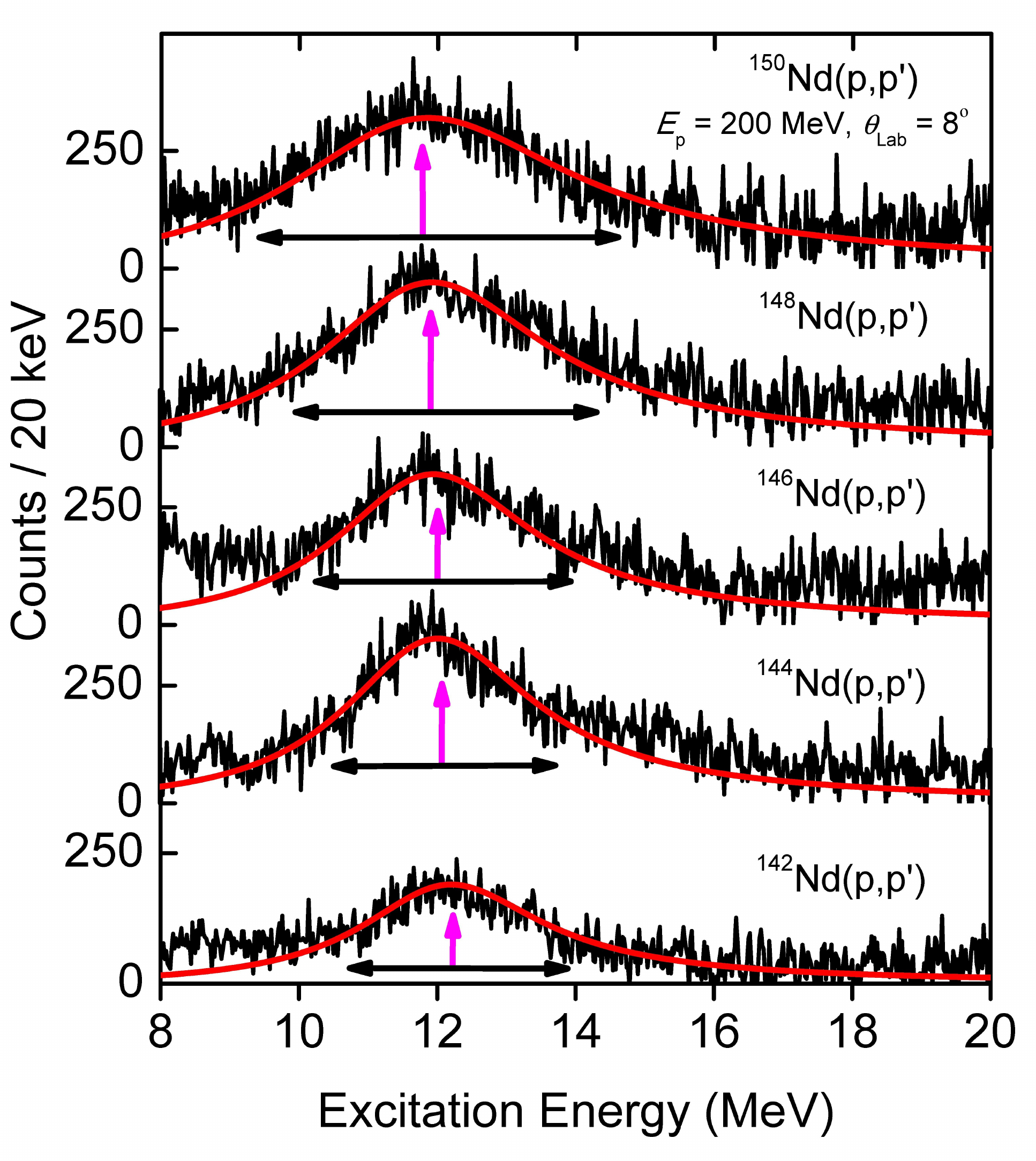}
\caption{Spectra from Fig.~\ref{fig:inelastic-all} after DWT background and $^{16}$O contaminant subtraction. 
The solid (red) lines are fitted Lorentzian function while the vertical (purple) and horizontal (black) arrows indicate the positions of the ISGQR centroids and the widths, respectively. 
The resonance width $\Gamma$ systematically increases from spherical to deformed nuclei.}
\label{fig:array-oxyfree}
\end{center}
\end{figure}

Figure~\ref{fig:inelastic-all} also shows that there is a sizable background under the ISGQR found between about 9 and 15 MeV.
This background stems from quasifree reactions and contributions from higher-$L$ and spinflip-type multipoles which are expected to have broad strength distributions.
Its shape and magnitude can be estimated from a decomposition of the spectra with a Discrete Wavelet Transform (DWT).
The method has been presented in detail in Refs.~\cite{Kal06,She08,Usm11b,Pol14} and further information on the wavelet analysis is given below.  
Final spectra obtained after both DWT and $^{16}$O contaminant subtraction (for details see \cite{Kur14}) are displayed in Fig.~\ref{fig:array-oxyfree}.  
The energy centroids of the ISGQR vary from 12.2 MeV in $^{142}$Nd to 11.9 MeV in $^{150}$Nd consistent with expectations from systematics \cite{Har01}.
The data are well described by single Lorentzians for all isotopes (red lines in Fig.~\ref{fig:array-oxyfree}) with a width increasing from 3.2 to 5.1 MeV when going from $^{142}$Nd to $^{150}$Nd.
 %
\section{Theoretical model calculations} \label{sec:model}
The evolution of giant resonance properties from spherical to deformed ground states in the stable Nd and Sm isotope chains has recently been studied in Ref.~\cite{Yos13} and more generally across the nuclear chart in Ref.~\cite{Sca14}.
The present calculations are performed within the Skyrme separable Random Phase Approximation approach \cite{Ne06}. 
The method is fully self-consistent since  both the mean field and residual interaction are derived from the same Skyrme functional.
The residual interaction includes all the functional contributions as well as the Coulomb-direct and -exchange terms. 
The self-consistent factorization of the residual interaction crucially reduces  the computational effort for deformed nuclei while maintaining the high accuracy of the calculations (see e.g.\ \cite{Ne06,Ne_IJMPE_08,kle_PRC_08}). 

The Skyrme parameterization SVmas10 \cite{Kl09} is used for the ISGQR description.
It was chosen because studies of different Skyrme forces \cite{Ne06,Ne_IJMPE_08} indicate a strong sensitivity of ISGQR properties on the effective mass $m^*$ and large values of $m^*/m$ provide the best results. 
The code exploits the 2D grid in cylindrical coordinates with the equilibrium axial quadrupole deformation determined by minimization of the total energy. 
In this work, the extracted deformation parameters  $\beta_2 = 0.135$, 0.192 and 0.285 for $^{146}$Nd, $^{148}$Nd and $^{150}$Nd, respectively, were found to be in good agreement with those deduced from experimental $B(E2)$ strengths exciting the first $2^+$ state \cite{Ram01}.
Pairing is treated with delta forces at the BCS level. 
In the residual interaction, the pairing particle-particle channel is taken into account. 
Deformation-induced coupling between quadrupole, hexadecapole and monopole modes is also embraced \cite{Ne06}. 
A large two-quasiparticle configuration space ($\sim$190000 states with energies up to 70 MeV) is taken into account, and the ISGQR energy-weighted sum rule is approximately exhausted for all isotopes. 
%
\begin{figure}[tbh!]
\begin{center}
\includegraphics[width=\columnwidth,angle=0]{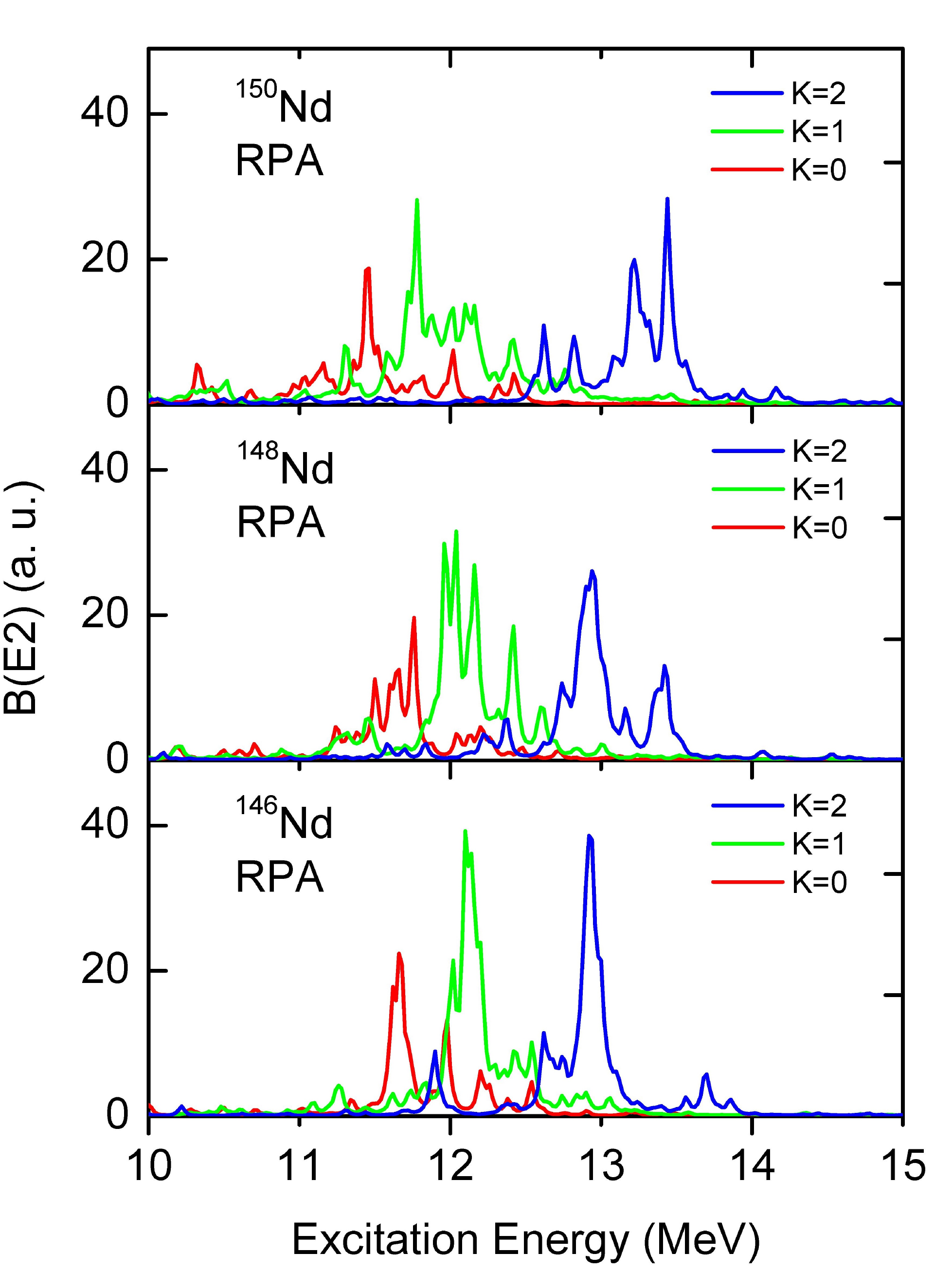}
\caption{Results for Skyrme SVmas10 RPA calculations for the isoscalar $E2$ strength functions of the deformed $^{146}$Nd, $^{148}$Nd and $^{150}$Nd nuclei. 
Splitting can be clearly observed between $K = 0$ (green) or $K$ = 1 (red) and $K$ = 2 (blue) components of the ISGQR, increasing with nuclear deformation.
}
\label{fig:rpa-array}
\end{center}
\end{figure}

Input for the wavelet analysis is provided by the SRPA strength function of the isoscalar $E2$  strength weighted by a Lorentzian with averaging parameter $d=45$ keV equal to the experimental resolution. 
The small value of the averaging allows to resolve the fine structure of the calculated ISGDR for the aims of the wavelet and semblance analysis described below.
The corresponding results for $^{146,148,150}$Nd are displayed in Fig.~\ref{fig:rpa-array}.
The strength functions are decomposed into the $K$ = 0 (red), 1 (green), 2 (blue) components with centroid energies 11.94 MeV, 12.21 MeV,  12.82 MeV ($^{146}$Nd), 11.75 MeV, 12.12 MeV, 12.92 MeV  ($^{148}$Nd) and 11.59 MeV, 12.02 MeV, 13.13 MeV ($^{150}$Nd).
One observes not only a fragmentation of the 1p-1h strength for a given $K$-value (Landau damping) but also a shift between the centroids ($K$-splitting) which increases with ground-state deformation.
While the effect is small for the splitting between $K = 0$ and 1 rising from 0.3 to 0.4 MeV, the difference between $K = 1$ and 2 is about 0.6 MeV in $^{146}$Nd increasing to about 1.1 MeV in $^{150}$Nd.  


\section{Wavelet analysis} \label{sec:wavelet}

Wavelet analysis has been established as a tool to quantitatively analyze the fine structure of nuclear giant resonances.
A detailed description of the methods and a critical comparison  with other approaches can be found in Ref.~\cite{She08}.
Here, we briefly review basics of the application to nuclear spectra. 
Wavelet analysis can be regarded as an extension of Fourier analysis which allows to conserve the correlation between the observable and its transform.
In the present case the coefficients of the wavelet transform are defined as
\begin{equation}
   C\left( {\delta E,E_{x}} \right) = \int\limits_{ - \infty
   }^\infty  {\sigma\left( E \right)\Psi \left( {\delta E,E_{x},E}
   \right)dE}.
   \label{eq:cwt}
\end{equation}
They depend on two parameters, a scale $\delta E$ stretching and compressing the wavelet $\Psi$($E$), and a position $E_{\rm x}$ shifting the wavelet in the spectrum $\sigma$($E$). 
The variation of the variables can be carried out with continuous or discrete steps. 
The analysis of the fine structure of giant resonances is performed using the continuous wavelet transform (CWT), where the fitting procedure can be adjusted to the required precision. 

In order to achieve an optimal representation of the signal using wavelet transformation, one has to select a wavelet function $\Psi$ which resembles the properties of the studied signal $\sigma$. 
A maximum of the wavelet coefficients at certain value $\delta$E indicates a correlation in the signal at the given scale, also called characteristic scale.
The best resolution for nuclear spectra is obtained with the so-called Complex Morlet wavelet (cf.\ Fig.~9 in Ref.~\cite{She08}) because the detector response closely resembles a Gaussian line shape and the Complex Morlet wavelet is a product of Gaussian and cosine functions
\begin{equation}
\Psi (x) = \frac{1}{\sqrt{\pi f_b}}{\rm exp}(2 \pi i f_c){\rm exp}\left(-\frac{x^2}{f_b}\right),
\label{eq:morlet}
\end{equation}
where $f_{\rm c}$ is the wavelength centre frequency and $f_{\rm b}$ is the bandwidth parameter. 
The scale values obtained from the Complex Morlet function are also consistent with those of Fourier analysis, but the wavelet analysis allows one to relate them to the energy region of the ISGQR. It should be noted that the wavelet coefficients obtained are complex.

Alternatively, a spectrum decomposition based on the DWT can be used, where scales and positions in the wavelet analysis are varied by powers of two. 
It allows an iterative decomposition of the spectrum by filtering it into two signals, approximations ($A$) and details ($D$), representing the large-scale (low-frequency) and small-scale (high-frequency) part for a given scale region analog to the effect of high- and low-pass filters in an electric circuit. 
In each step $i$ of the decomposition, the initial signal $\sigma$($E$) can be reconstructed as
\begin{equation}
\sigma(E)=A_i + \sum D_i.
\label{eq:dwtapp+det}
\end{equation}
This operation can be repeated until the individual detail consists of a single bin. 

A DWT can only be performed with wavelets which possess a so-called scaling function \cite{She08}.
This is not the case for the Complex Morlet wavelet, thus the Bior6.8 wavelet function \cite{Mal99} is used as an alternative.  
It provides another useful property for a determination of background in the data. 
Each wavelet function can be characterized by its number of vanishing moments,
\begin{equation}
   \label{eq:vanishingm}
   \int\limits_{ - \infty }^\infty  {E^n \Psi \left( E \right)dE =
   0,\;\; n = 0,1...m}.
\end{equation}
For Bior6.8 the number is equal to six, i.e. any background in the spectrum that can be approximated by a polynomial function up to order five does not contribute to the wavelet coefficients. 
Such an analysis was performed for the spectra of Fig.~\ref{fig:inelastic-all}.
One can identifiy the decomposition order $i$ containing the largest scale, i.e.\ the resonance width \cite{Kal06,She08,Usm11b,Pol14}.
The next-higher order provides the form of the background.
For the subtraction (cf.\ Fig.~\ref{fig:array-oxyfree}) the resulting functions were normalized to achieve background-free spectra between the proton and neutron thresholds.
   
\begin{figure}[t]
\begin{center}
\includegraphics[width=\columnwidth,angle=0]{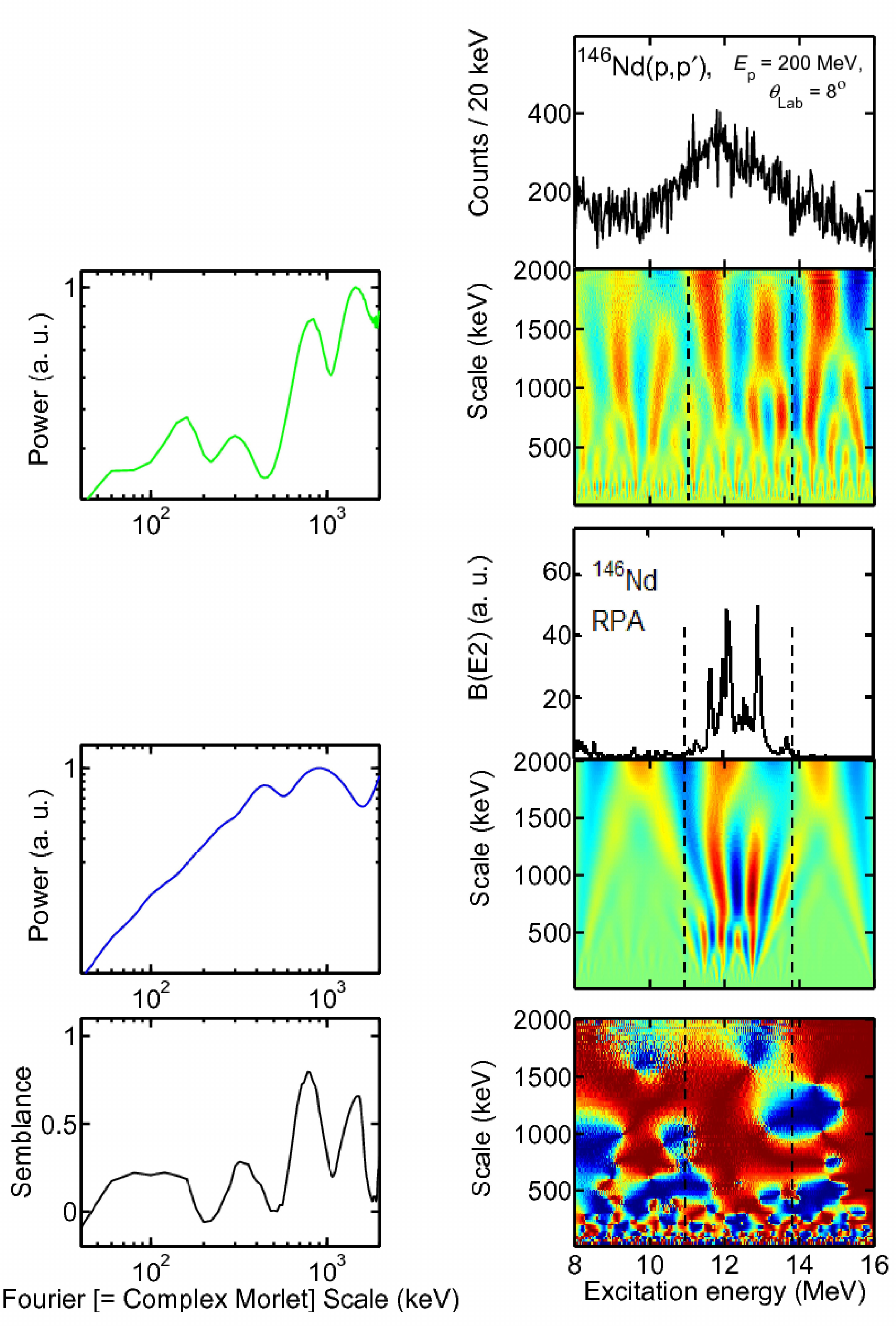}
\caption{Wavelet analysis for $^{146}$Nd.
$Top$: Background-subtracted spectrum of the ISGQR in the $^{146}$Nd(p,p$'$) reaction and distribution of wavelet transforms, Eq.~(\ref{eq:cwt}), (r.h.s.) and wavelet power spectrum (l.h.s).
$Middle$: Same for theoretical B($E$2) strength distribution.
$Bottom$: Semblance, Eq.~(\ref{eq:semblance}), (r.h.s) and semblance power spectrum (l.h.s.) integrated over the energy range indicated by the vertical dashed lines. 
}
\label{fig:semb-nd146}
\end{center}
\end{figure}
Figure~\ref{fig:semb-nd146} illustrates the application of the CWT on both experimental and theoretical results for the case of  $^{146}$Nd. 
The top panels on the r.h.s.\ present the experimental spectrum and a 2D distribution of the real part of the wavelet coefficients according to Eqs.~(\ref{eq:cwt}, \ref{eq:morlet}), where the color code indicates large positive (red) or negative (blue) coefficients and yellow corresponds to small values of $C(\delta E,E_{\rm x})$.
Projection on the scale axis leads to the power spectrum shown on the l.h.s., summed between the vertical dashed lines defined by the width of the experimental ISGQR strength function.
One observes maxima in the power spectrum which are characteristic scales of the fluctuations in the analyzed spectrum.   
The middle part of Fig.~\ref{fig:semb-nd146} presents an analoguous analysis of the theoretically predicted $E2$ strength function from Sect.~\ref{sec:model}.
Although no scales are observed on the theoretical power spectrum for smaller scale values, pronounced characteristic scales are visible around 1 MeV resembling the experimental results. 
These are  most likely caused by the splitting between the main fragments of $K$ = 0, 1 and 2 strengths as demonstrated in the next Section.

\section{Wavelet signature for $K$-splitting from a semblance analysis} \label{sec:semblance}

A quantitative measure of the correspondence between the two sets of wavelet coefficents is provided by a wavelet-based semblance analysis, where the local phase relationships of the complex wavelet coefficients can be studied as a function of scale \cite{Coo08}.
The semblance $S$ can be expressed as
\begin{equation}
\label{eq:semblance}
S = \cos^n (\theta)~,
\end{equation}
where $n$ is an odd integer greater than zero ($n = 1$ in the present case), yielding values ranging from -1 (inversely correlated) through zero (uncorrelated) to +1 (correlated).   
Here, the local phase $\theta$ is given by $\theta = \tan^{-1} [\Im(C_{1,2})/\Re(C_{1,2})]$, where the cross-coefficient $C_{1,2} = C_1 C^*_2$ with $C_1$ the wavelet transform of data set 1 and $C^*_2$ the complex conjugate of dataset 2. 
To the best of our knowledge, the following represents the first application of semblance analysis to a nuclear structure problem.

The bottom part of Fig.~\ref{fig:semb-nd146} shows the result from the application of Eq.~(\ref{eq:semblance}) to the experimental spectrum and the RPA prediction.
Since the accuracy of giant resonance energies predicted in mean-field models is on the level of a few hundred keV only, a shift between the experimental and theoretical spectrum was allowed for. 
Optimum semblance values were achieved for a 200 keV downward shift of the RPA result with respect  to the data.
Because of the phase, the semblance (r.h.s.) can have positive (red) or negative (blue) values.
A large positive correlation is obtained over most of the resonance  -- in this case between $E_{\rm x}$ = 11 to 13 MeV where the RPA $E2$ strength lies -- for scale values corresponding to two characteristic scales around 1 MeV.
The l.h.s. is again the power spectrum summed over the energy region indicated by the vertical dashed lines showing a near maximum correlation for the  two characteristic scales produced by the splitting between the main $K$ = 1 and 2 fragments as well as between the $K$ = 0 peak (plus the $K$ =1 strength fragment below 12 MeV) and the main $K$ = 2 fragment in the RPA calculations.     

For smaller scale values the semblance shows large fluctuations from correlation to anti-correlation over the energy region of the resonance.
The theoretical power spectrum indicates no characteristic scale below 500 keV while the data show scales around 150 and 300 keV. 
This points towards an origin of these scales from coupling to 2p-2h states \cite{She04,She09} which is outside of the RPA model space.


Further evidence for the interpretation of characteristic scales to arise from $K$-splitting is provided by a semblance analysis of  $^{148}$Nd and $^{150}$Nd, respectively. 
For $^{148}$Nd the RPA strength distribution shows very similar pattern to $^{146}$Nd and indeed two scales  are also found which can be identified with the energy shift of the main parts of the $K$ = 0 and 1 strength relative to the $K$ = 2 strength.
%
\begin{figure}[t]
\begin{center}
\includegraphics[width=\columnwidth,angle=0]{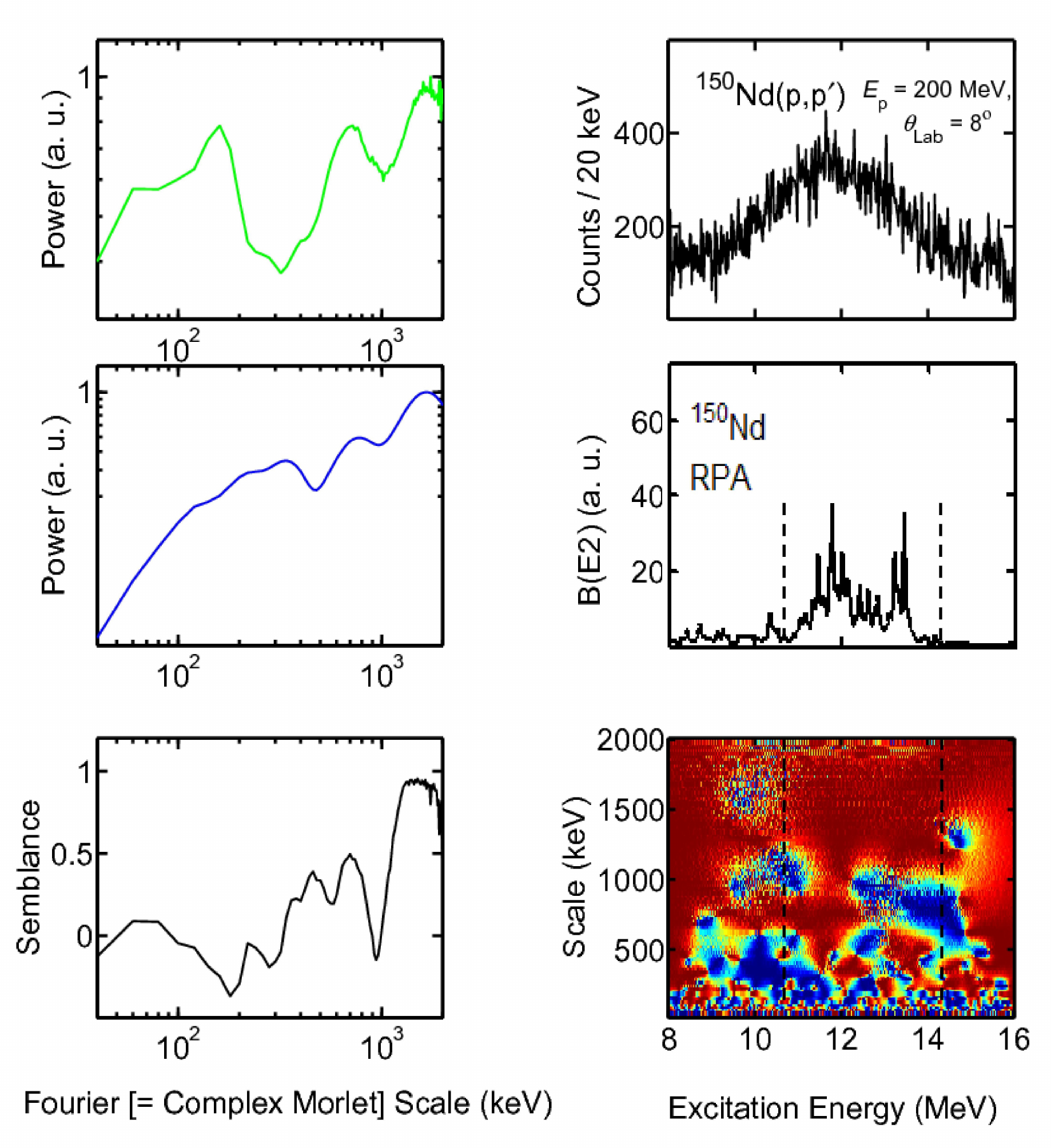}
\caption{Wavelet analysis for $^{150}$Nd.
$Top$: Background-subtracted spectrum of the ISGQR in the $^{150}$Nd(p,p$'$) reaction (r.h.s.) and wavelet power spectrum (l.h.s).
$Middle$: Same for theoretical B($E$2) strength distribution.
$Bottom$: Semblance, Eq.~(\ref{eq:semblance}), (r.h.s) and semblance power spectrum (l.h.s.) integrated over the energy range indicated by the vertical dashed lines.  
}
\label{fig:semb-nd150}
\end{center}
\end{figure}
%
Because of the increasing deformation the theoretical ISGQR strength distribution in $^{150}$Nd becomes more fragmented and the energy shift between the $K$ components is significantly larger than in $^{146,148}$Nd.
This behavior is again well reproduced in the semblance analysis (bottom l.h.s.\ of Fig.~\ref{fig:semb-nd150}).
Because of the broadening of the peak in the RPA strength distributions at lower excitation energies the corresponding semblance analysis also shows one broad correlation maximum shifted overall to a higher scale value compared with $^{146,148}$Nd.
The theoretically expected increase of $K$-splitting with ground-state deformation seems to be  reflected in the properties of the experimental fine structure. 

A remaining question is the influence of the chosen interaction on the semblance results.
As pointed out in Section \ref{sec:model}, the bulk parameter most sensitive to the ISGQR is the effective mass and values $m^*/m \approx 1$ provide the best systematic description of the ISGQR energy centroid. 
Therefore, we repeated the analysis using some widely used Skyrme functionals (SVbas, SkM* and Sly6) with varying $m^*/m$ values (0.9, 0.79, 0.69). 
For $^{146}$Nd, the additional interactions produce similar E2 strength distributions, but shifted up in excitation energy by up to 2 MeV with decreasing effective mass, inconsistent with the experimental data.
For $^{150}$Nd, due to the larger  $\beta_2$ deformation parameters they lead to a broadening of the $E2$ strength distributions with respect to SVmas10 and the experimental results.
 
\begin{figure}[t]
\begin{center}
\includegraphics[width=0.7\columnwidth,angle=0]{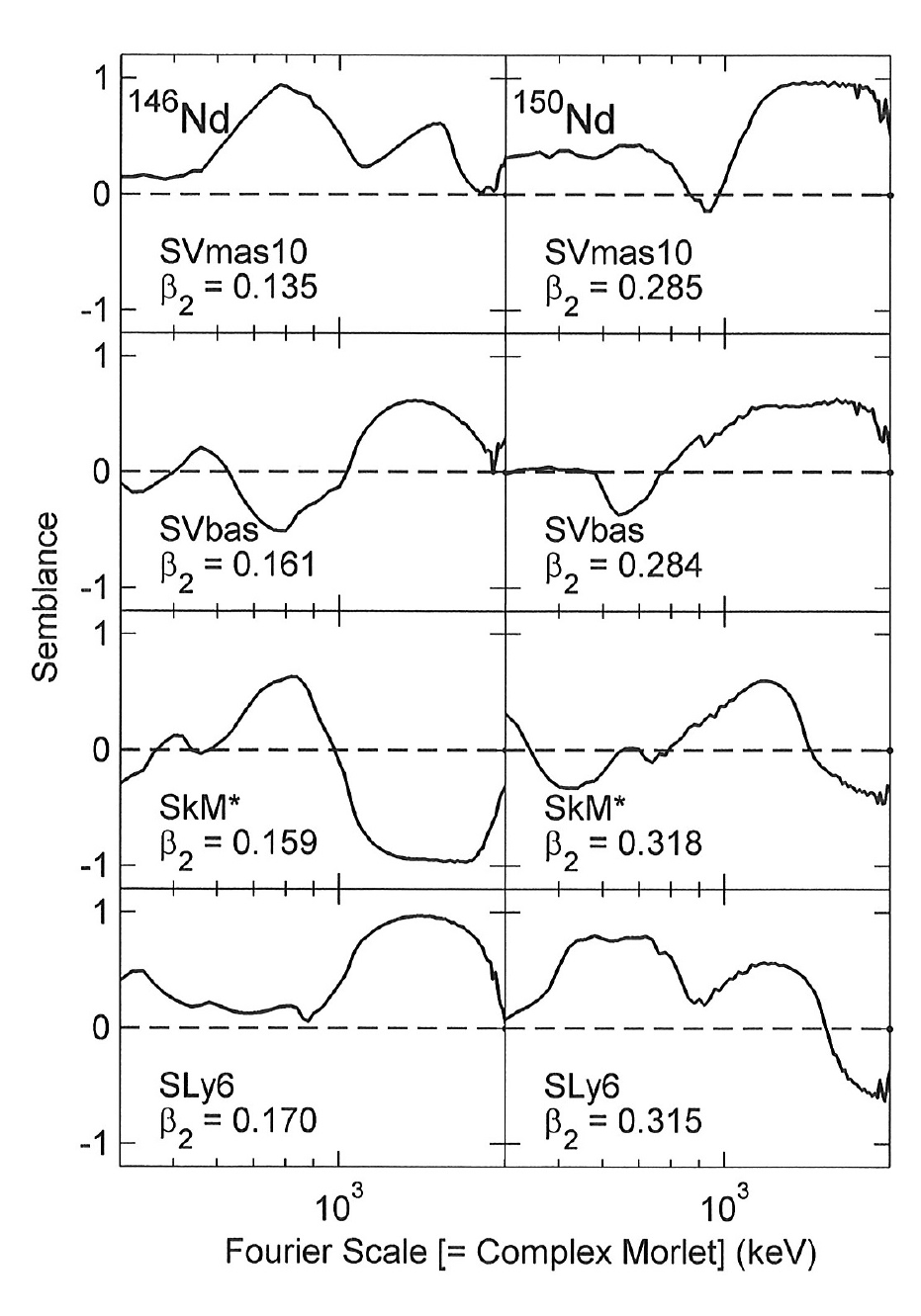}
\caption{Comparison of semblance analysis for different Skyrme interactions (see text).
Left: $^{146}$Nd.
Right: $^{150}$Nd.
}
\label{fig:semblance-comp}
\end{center}
\end{figure}

The corresponding semblances together with those obtained with SVmas10 (cf.\ Figs.~\ref{fig:semb-nd146}, \ref{fig:semb-nd150}) are displayed in Fig.~\ref{fig:semblance-comp} for $^{146}$Nd and $^{150}$Nd, respectively. 
The scale axis is limited to values $> 400$ keV, where a comparison of experimental and theoretical results is meaningful, see above.
Figure \ref{fig:semblance-comp} clearly demonstrates the sensitivity of the semblance observable to the chosen interaction. 
None of the other interactions can reproduce the semblance peaks in $^{146}$Nd obtained with SVmas10 consistent with the wavelet scales.
Rather, one scale is missing (SLy6) or even strong anti-correlation (SVbas, SkM*) is found.
For $^{150}$Nd, SVbas roughly reproduces the broad maximum, although with reduced correlation with respect to SVmas10, while SkM* and Sly6 again show anti-correlation instead of correlation for larger scale values.     
%
%
\section{Conclusions}\label{sec:conclusions}

The present work provides evidence for the phenomenon of fine structure of the ISGQR in the quadrupole-deformed $^{146,148,150}$Nd isotopes.
The observation of fine structure in heavy deformed nuclei is far from trivial considering the extremely high level densities of $2^+$ states in the corresponding excitation energy region ($10^7 - 10^8$ MeV$^{-1}$ estimated from Refs.~\cite{rau97,gor08}).
It requires that the effect of coupling of the doorway states (the main fragments of the 1p-1h strength) to the surrounding more complex states is sufficiently weak.

It is shown by comparison with Skyrme RPA calculations based on the SVmas10 interaction that characteristic scales of a wavelet analysis describing the experimental fine structure can be traced back to the energy splitting between $K$ components of the isoscalar $E$2 strength in deformed nuclei.
Quantitative evidence is provided by a semblance analysis which considers phase correlations of the wavelet transforms of experimental and theoretical spectra at a given scale.
Remarkably, not only the experimental strength functions can provide information on the $K$-splitting from the observation of overall larger widths in deformed nuclei but also the fine structure carries a clear signature.
The comparison of results from different interactions illustrates the sensitivity of wavelet observables, which may help to improve the development of energy density functionals capable of accounting for deformation degrees of freedom.

\section*{Acknowledgements}
We are indebted to J.L.\ Conradie and the accelerator crew at iThemba LABS for providing excellent beams.
This work has been supported by the South African National Research Foundation and by the DFG under contract SFB 1245. 
J.K.\ acknowledges the Czech Science Foundation for support under project P203-13-07117S.

\section*{References}

\end{document}